\documentclass{singlecol-new}

\usepackage{graphicx}
\usepackage{amsmath}
\usepackage{natbib}

\theoremstyle{TH}{

}

\theoremstyle{THhit}{

}

\theoremstyle{THrm}{

}

\makeatletter
\def\theequation{\arabic{equation}}
\makeatother

\begin{document}

\setcounter{page}{651}

\LRH{D. Setsirichok et al.}

\RRH{Small ancestry informative marker panels}

\VOL{6}

\ISSUE{6}

\BottomCatch

\PAGES{2012}

\CLline

\PUBYEAR{2012}

\subtitle{}

\title{Small ancestry informative marker panels for complete classification between the original four HapMap populations}

\authorA{Damrongrit Setsirichok\\ and Theera Piroonratana}

\affA{Department of Electrical Engineering,\\ Faculty of Engineering,\\
King Mongkut's University of Technology North Bangkok,\\
1518 Piboolsongkram Road,\\ Bangsue, Bangkok 10800, Thailand\\
E-mail: d.setsirichok@gmail.com\\
E-mail: theepi@gmail.com}

\authorB{Anunchai Assawamakin\vs{-3}}

\affB{Biostatistics and Informatics Laboratory,\\ Genome Institute,\\
National Center for Genetic Engineering and Biotechnology,\\
National Science and Technology Development Agency,\\
113 Thailand Science Park,\\ Phahonyothin Road,\\
Klong 1, Klong Luang, Pathumthani 12120, Thailand\\
Email: anunchai\_ice@yahoo.com}

\authorC{Touchpong Usavanarong\vs{-3}}

\affC{Department of Electrical Engineering,\\ Faculty of Engineering,\\
King Mongkut's University of Technology North Bangkok,\\
1518 Piboolsongkram Road,\\ Bangsue, Bangkok 10800, Thailand\\
E-mail: blood\_serpent@hotmail.com}

\authorD{Chanin Limwongse\vs{-3}}

\affD{Division of Molecular Genetics,\\ 
Department of Research and Development,\\
Faculty of Medicine Siriraj Hospital,\\ Mahidol University,\\
2 Prannok Road,\\ Bangkoknoi, Bangkok 10700, Thailand\\
Email: siclw@mahidol.ac.th}

\authorE{Waranyu Wongseree\vs{-3}}

\affE{Department of Electrical Engineering,\\ Faculty of Engineering,\\
King Mongkut's University of Technology North Bangkok,\\
1518 Piboolsongkram Road,\\ Bangsue, Bangkok 10800, Thailand\\
E-mail: waranyu.wongseree@gmail.com}

\authorF{Chatchawit Aporntewan\vs{-3}}

\affF{Department of Mathematics,\\ Faculty of Science,\\
Chulalongkorn University,\\
254 Phayathai Road,\\ Pathumwan, Bangkok 10330, Thailand\\
Email: chatchawit.a@chula.ac.th}

\authorG{Nachol Chaiyaratana*\vs{-3}}

\affG{Department of Electrical Engineering,\\ Faculty of Engineering,\\
King Mongkut's University of Technology North Bangkok,\\
1518 Piboolsongkram Road,\\ Bangsue, Bangkok 10800, Thailand\\
and\\
Division of Molecular Genetics,\\ 
Department of Research and Development,\\
Faculty of Medicine Siriraj Hospital,\\ Mahidol University,\\
2 Prannok Road,\\ Bangkoknoi, Bangkok 10700, Thailand\\
E-mail: nchl@kmutnb.ac.th\\
E-mail: n.chaiyaratana@gmail.com\\
{*}Corresponding author}

\begin{abstract}
A protocol for the identification of ancestry informative markers (AIMs) from genome-wide single nucleotide polymorphism (SNP) data is proposed. The protocol consists of three main steps: (a)~identification of potential positive selection regions via $F_{ST}$ extremity measurement, (b) SNP screening via two-stage attribute selection and (c) classification model construction using a na\"ive Bayes classifier. The two-stage attribute selection is composed of a newly developed round robin symmetrical uncertainty ranking technique and a wrapper embedded with a na\"ive Bayes classifier. The protocol has been applied to the HapMap Phase II data. Two AIM panels, which consist of 10 and 16 SNPs that lead to complete classification between CEU, CHB, JPT and YRI populations, are identified. Moreover, the panels are at least four times smaller than those reported in previous studies. The results suggest that the protocol could be useful in a scenario involving a larger number of populations. 
\end{abstract}

\KEYWORD{AIM; Ancestry informative marker; Attribute selection; $F_{ST}$; HapMap; Heterozygosity; Population classification; Positive selection; SNP; Single nucleotide polymorphism.}

\REF{to this paper should be made as follows: Setsirichok, D., Piroonratana, T., Assawamakin, A., Usavanarong, T., Limwongse, C., Wongseree, W., Aporntewan, C. and Chaiyaratana, N. (2012) `Small ancestry informative marker panels for complete classification between the original four HapMap populations', {\it International Journal of Data Mining and Bioinformatics}, Vol.~6, No.~6, \h{pp.651--674}.}

\begin{bio}
Damrongrit Setsirichok is a Ph.D. student at the Department of Electrical Engineering, Faculty of Engineering, King Mongkut's University of Technology North Bangkok. He also received his B.Eng. degree in computer engineering from King Mongkut's University of Technology North Bangkok. His research interests include machine learning, bioinformatics and genetic epidemiology.\vs{9}

\noindent Theera Piroonratana is a post-doctoral researcher at the Department of Electrical Engineering, Faculty of Engineering, King Mongkut's University of Technology North Bangkok. He also received his B.Eng. and M.Eng. degrees in production engineering as well as his Ph.D. degree in electrical engineering from King Mongkut's University of Technology North Bangkok. His research interests include evolutionary multi-objective optimisation and machine learning.\vs{9}

\noindent Anunchai Assawamakin is a recipient of the Post-doctoral Fellowships Programme at the National Science and Technology Development Agency, offered through the National Center for Genetic Engineering and Biotechnology. He received his B.Sc. degree in pharmacy and his Ph.D. degree in immunology from Mahidol University. His research interests include human genetics, genetic epidemiology, population genetics and bioinformatics.\vs{9}

\noindent Touchpong Usavanarong is an M.Eng. student at the Department of Electrical Engineering, Faculty of Engineering, King Mongkut's University of Technology North Bangkok. He also received his B.Eng. degree in electrical engineering from King Mongkut's University of Technology North Bangkok. His research interests include machine learning, bioinformatics and genetic epidemiology.\vs{9}

\noindent Chanin Limwongse is the Head of Division of Molecular Genetics at the Department of Research and Development, Faculty of Medicine Siriraj Hospital, Mahidol University. He also received his M.D. degree from Mahidol University. His research interests include human genetics and genetic diseases.\vs{9}

\noindent Waranyu Wongseree is a post-doctoral researcher at the Department of Electrical Engineering, Faculty of Engineering, King Mongkut's University of Technology North Bangkok. He also received his B.Eng., M.Eng. and Ph.D. degrees in electrical engineering from King Mongkut's University of Technology North Bangkok. His research interests include machine learning, evolutionary computation and bioinformatics.\vs{9}

\noindent Chatchawit Aporntewan is an assistant professor of computer science at the Department of Mathematics, Chulalongkorn University. He also received his B.Eng., M.Eng. and Ph.D. degrees in computer engineering from Chulalongkorn University. His research interests include machine learning, evolutionary computation and bioinformatics.\vs{9}

\noindent Nachol Chaiyaratana is an associate professor of electrical engineering at King Mongkut's University of Technology North Bangkok and an adjunct professor of genetic epidemiology at Mahidol University. He received his B.Eng. and Ph.D. degrees from the Department of Automatic Control and Systems Engineering, University of Sheffield. His research interests include evolutionary computation, machine learning and genetic epidemiology.
 \end{bio}

\maketitle

\section{Introduction}
Human evolution can be traced via various forms of genetic information~\citep{Quintana, Jobling01, Jobling03, Mitrofanova}. Many studies involving mitochondrial DNA (mtDNA)~\citep{Quintana, Salas, Metspalu} and DNA variation on the Y chromosome~\citep{Jobling03, Karafet} reveal that the present human species is originated from Africa~\citep{Lewin}. In fact, the migration of behaviourally modern humans from Africa to all continents takes place only approximately 70,000--50,000 years ago~\citep{Mellars06}. Through this course of migration, the population subdivision has occurred and has resulted in the emergence of new populations and ethnic groups.

\enlargethispage{-1.3pc}

With the presence of strong evidence that supports the occurrence of population subdivision, the genetic description of a population can be established. Furthermore, the clustering of individuals into many populations with different genetic backgrounds can be done automatically~\citep{Pritchard, Falush03, Tang, Falush07, Gao, Paschou}. The task of assigning an unknown individual to the correct population can be carried out by inspecting his or her population-specific genetic patterns once the population boundary is defined via genetics or self-reported ethnicity. These patterns usually consist of ancestry informative markers (AIMs)---genetic markers that exhibit substantially different allele frequencies between populations of descendants derived from mutually inbred ancestors. The identification of AIMs has been proven to be beneficial to many research areas including genetic epidemiology~\citep{Enoch, Seldin, Tian, Baye} and forensic science~\citep{Phillips, Budowle}.

The international HapMap project discovers over 3,000,000 single nucleotide polymorphisms (SNPs) in the genome of each human individual~\citep{HapMap03, HapMap05, HapMap07}. As a result, the search for SNP-based AIMs usually involves genome-wide SNP screening. Many measures including informativeness~\citep{Rosenberg03, Rosenberg05}, {\it t} statistics~\citep{Park, Zhou} and {\it F}~statistics~\citep{Zhou} have been proposed for SNP prioritisation. The screening is then carried out via a greedy search~\citep{Rosenberg05} or a ranking method~\citep{Zhou}. Once SNPs have been selected, their capability as AIMs can be validated via classification model construction. The classification task specifically involves the use of genotypic attributes from selected SNPs as inputs for identifying the ethnicity or population label of an individual. Standard machine learning techniques that have been successfully implemented as classifiers include a support vector machine~\citep{Zhou} and genetic programming~\citep{Nunkesser}. The same two-step protocol, which involves SNP screening and classification model construction, has also been successfully applied to genetic association studies~\citep{Moore}.

Genome-wide SNP screening indicates that AIMs extracted from the HapMap data spread across the whole genome~\citep{Park, Paschou, Zhou}. In fact, only 14 SNPs are required for the complete classification between three populations namely the CEU (Utah residents with northern and western European ancestry) population, the YRI (Yoruba in Ibadan, Nigeria) population and the Asian population obtained by merging the JPT (Japanese in Tokyo) and CHB (Han Chinese in Beijing) populations together~\citep{Paschou}. However, 64 SNPs are needed for the near complete classification between all four HapMap populations, indicating that additional 50 SNPs are required for the classification between CHB and JPT populations~\citep{Paschou}. This implies that large AIM panels are necessary when the classification task involves multiple populations which are closely related to one another. In order to make the AIM identification task tractable for this kind of scenario, it is crucial to develop a protocol that leads to the discovery of the smallest possible AIM panels.

Early works on AIM identification are usually conducted by exploiting little prior knowledge regarding population subdivision. By incorporating the prior knowledge into the AIM search protocol, it is possible that the search can be limited to specific genomic regions. The regions that are strong candidates for this consideration are positive selection regions~\citep{Olson, Sabeti02, Bamshad, Akey, Vallender, Voight, Sabeti07}. This is because one of the main signatures of positive selection is the decrease of heterozygosity over Hardy-Weinberg expectations~\citep{Beaumont}, which also signifies population subdivision. The search for positive selection has been conducted on samples from many populations including European, African and Asian~\citep{Hinds, Myles, Oleksyk, Pickrell}. The discovered selective regions spread across the whole genome and cover genes that govern growth, pigmentation, immune defence, carbohydrate metabolism, behaviour and other functions. 

It has been suggested that SNPs from positive selection regions can be used as AIMs~\citep{Lao, Phillips, Seldin, Tian}. This is because an AIM from a positive selection region detected in a multiple population data set has a strong potential for being directly applicable as an AIM for other data sets containing similar populations. Evidence that supports this suggestion includes the selection of a SNP from {\it EDAR} as a member of an AIM panel for inferring ancestors of many common populations in the US~\citep{Kosoy}. This gene involves in the development of hair follicles and has undergone positive selection in Asian populations~\citep{Sabeti07, Bryk, Fujimoto, Mou}. Nonetheless, an attempt to extract entire AIM panels from positive selection regions has never been made. 

In this article, a protocol for identifying AIMs from potential positive selection regions is proposed. It is aimed that by concentrating the AIM search on potential positive selection regions, the resulting AIM panels should be smaller than those identified without the genomic region restriction. The proposed protocol involves three main steps: identification of SNPs in potential positive selection regions, SNP screening via attribute selection and classification model construction. Potential positive selection regions are located by means of $F_{ST}$ extremity measurement~\citep{Bamshad, Sabeti07, Bryk, Fujimoto, Myles}. SNPs with extreme $F_{ST}$ values are subsequently screened by the most appropriate technique selected from a number of filter- and wrapper-based attribute selection techniques~\citep{Saeys, Zheng} including a correlation-based feature selection technique~\citep{Hall}, a wrapper embedded with a na\"ive Bayes classifier~\citep{Kohavi}, a simple symmetrical uncertainty~\citep{Press} ranking technique and a newly proposed round robin symmetrical uncertainty ranking technique. Finally, the classification model is constructed by a na\"ive Bayes classifier. The functionality of the proposed protocol is demonstrated via an application to the HapMap data.

\section{Materials and methods}
\subsection{Data set for AIM identification}
The data set explored in this study is obtained from the public release \#23a of HapMap data set (Phase II, release date: March 2008), which is available in NCBI build 36 (dbSNP b126) coordinates. The data set consists of 3,619,209 SNPs in which the genotypic attribute value according to each SNP can be a homozygous wild-type, heterozygous or homozygous mutant genotype. These SNPs are extracted from 270 samples representing four populations: CEU, CHB, JPT and YRI. Both CEU and YRI data sets consist of 90 related samples---30 father-mother-offspring trios. In contrast, both CHB and JPT data sets contain 45 unrelated samples. Since the original HapMap data set is composed of related and unrelated samples, only 210 unrelated samples are considered. The sample reduction is carried out by removing offspring samples from both CEU and YRI data sets.

\subsection{$F_{ST}$ extremity measurement}
The decrease of heterozygosity over Hardy-Weinberg expectations due to population subdivision can be described by an $F_{ST}$ measure \citep{Wright}. $F_{ST}$ is defined by
%
%
\begin{equation}
F_{ST} = \frac{H_T - H_S}{H_T}
\label{eq:Fst}
\end{equation}
where $H_S$ is the average of expected heterozygosities over all populations and $H_T$ is the expected heterozygosity in the combined population. $H_S$ is given by
%
%
\begin{equation}
H_S = \sum_i d_i (2p_i (1 - p_i))
\label{eq:Hs}
\end{equation}
where $d_i$ is the proportion of the {\it i}th population in the combined population and $p_i$ is the major allele frequency of the {\it i}th population. Similarly, $H_T$ is denoted by
%
%
\begin{equation}
H_T  = 2\bar p(1 - \bar p)
\label{eq:Ht}
\end{equation}
where $\bar p$ is the average of $p_i$ over all populations and is equal to $\sum_i d_i p_i$. Since population subdivision causes a perceived deficiency of heterozygotes, an $F_{ST}$ value is always between zero and one.

The search for SNPs with extreme $F_{ST}$ values is proven to be useful for preliminary screening for positive selection~\citep{Bamshad, Sabeti07, Bryk, Fujimoto, Myles}. An empirical distribution of $F_{ST}$ is first estimated from either some or all of available SNPs in the recruited samples~\citep{Sabeti07, Fujimoto, Myles}. The $F_{ST}$ extremity of each SNP is subsequently defined in terms of the percentile from the distribution. In this study, the $F_{ST}$ distribution is calculated for every population pair using all SNPs in the HapMap data.

\subsection{Attribute selection techniques}

\subsubsection{Simple symmetrical uncertainty ranking}
Symmetrical uncertainty is an information-theoretic measure discussed by \cite{Press}. Consider a classification problem that involves a sample set in which each sample is described by $n$ discrete-valued attributes (SNPs) and a class (population) label. Let $A$ be an attribute and $C$ be the class. The entropy $H$ of the class before and after observing the attribute is given by
%
%
\begin{equation}
H(C) = -\sum_{c \in C} p(c)\log_2 p(c)
\label{eq:cl_ent_before}
\end{equation}
and
%
%
\begin{equation}
H(C|A) = -\sum_{a \in A} p(a)\sum_{c \in C} p(c|a)\log_2 p(c|a),
\label{eq:cl_ent_after}
\end{equation}
respectively where $p$ denotes the probability value as estimated from the sample set. The difference between the entropy of the class before and after observing the attribute is the information gain~\citep{Quinlan} which is given by
%
%
{\setlength\arraycolsep{2pt}
\begin{eqnarray}
\mathit{Information} \mathit{Gain} & = & H(C) - H(C|A) \nonumber \\ 
                                   & = & H(A) - H(A|C) \nonumber \\ 
                                   & = & H(A) + H(C) - H(A,C).
\label{eq:info_gain}
\end{eqnarray}}The degree of correlation between the attribute and the class can subsequently be estimated via symmetrical uncertainty (SU) which is defined by
%
%
{\setlength\arraycolsep{2pt}
\begin{eqnarray}
SU & = & 2 \times \left[ \frac{H(A) + H(C) - H(A,C)}{H(A) + H(C)} \right] \nonumber \\
   & = & 2 \times \left[ \frac{H(C) - H(C|A)}{H(A) + H(C)} \right].
\label{eq:su}
\end{eqnarray}}It is noticeable that symmetrical uncertainty can be calculated from a quotient between the information gain and the sum of class entropy and attribute entropy. An attribute that has a high $SU$ value is highly correlated with the class and is also an important attribute for classification. A rank can be assigned to each attribute according to its $SU$ value where selected attributes are simply the top $n_r$ attributes with the highest ranks.

\subsubsection{Round robin symmetrical uncertainty ranking}
\begin{figure}[t!]
\centering
\caption{Outline of the SU$_2$ ranking. In this example, the three-population problem consists of balanced 150 samples and 1,000 SNPs. The genotype distribution of SNP$_1$ in all three populations is displayed. This leads to the $SU_2$ values of 0.016193, 0.009468 and 0.049025 for the population pairs (Pop$_1$, Pop$_2$), (Pop$_1$, Pop$_3$) and (Pop$_2$, Pop$_3$), respectively. After the calculation of $SU_2$ values for each SNP in every population pair is completed, SNPs are sorted according to their ranks. Three sets of top-ranked SNPs can be extracted from three population pairs. Only the top 50 SNPs are selected for each sorted set. The merging of three 50-SNP sets leads to the screened SNP set of size between 50 and 150.}
\includegraphics[width=12.5cm]{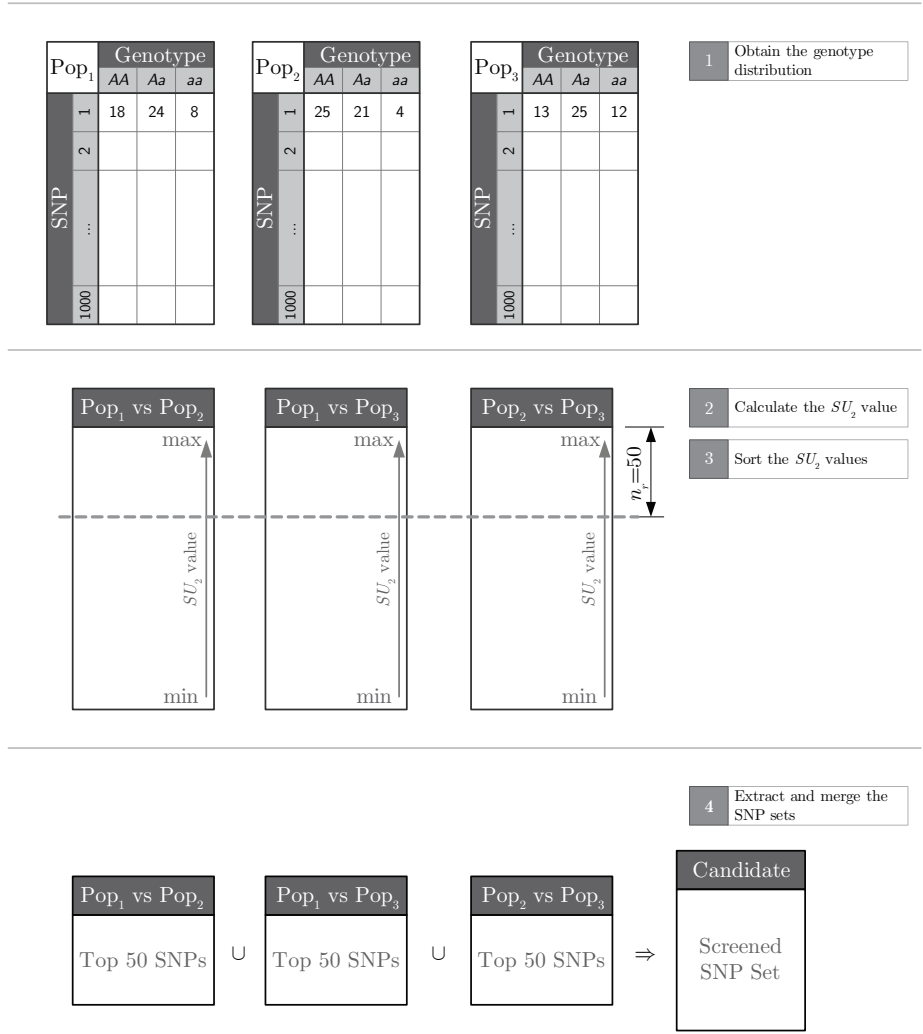}\\
\label{fig:SU2}
\end{figure}

From the previous section, it is noticed that $SU$ can be directly measured in classification problems with the number of classes greater than or equal to two. However, the calculation of $SU$ for a common SNP from two populations at a time is more useful for the identification of AIMs that are located in potential positive selection regions. This is because positive selection is generally confirmed when at least one new population is emerged from the ancestral population. For clarification, the measure is referred to as $SU_2$ when $SU$ is evaluated to determine the suitability of using an attribute for the classification between two classes. After the $SU_2$ values have been derived from all attributes, a rank can be assigned to each attribute; high $SU_2$ values lead to high ranks. The top $n_r$ attributes with the highest ranks are subsequently selected as screened attributes. For a multi-class problem, $\binom{n_c}{2} = n_c!/((n_c-2)!2!)$ sets of top-ranked attributes can be extracted from the data where $n_c$ is the number of classes. The merging of top-ranked attribute sets is subsequently carried out where the size of the merged attribute set is between $n_r$ and $n_r \times \binom{n_c}{2}$. The summary of round robin symmetrical uncertainty ranking (SU$_2$ ranking) is illustrated in Figure~\ref{fig:SU2}.

%
%

\subsubsection{Correlation-based feature selection technique}
A correlation-based feature selection technique~\citep{Hall} is an attribute subset evaluation heuristic that considers both the usefulness of individual features (attributes) in the classification task and the level of correlation among features. Each attribute subset is assigned a score given by
%
%
\begin{equation}
\mathit{Merit}_F = \frac{n_f \overline r_{\mathit{cf}}}{\sqrt{n_f + n_f(n_f-1) \overline r_{\mathit{ff}}}}
\label{eq:merit}
\end{equation}
where $\mathit{Merit}_F$ is the heuristic merit of an $n_f$-attribute subset $F$, $\overline r_{\mathit{cf}}$ is the average feature-class correlation and $\overline r_{\mathit{ff}}$ is the average feature-feature correlation. The correlation is obtained from the $SU$ measure. An attribute subset receives a high merit score if it contains features that are highly correlated with the class and at the same time have low correlation among one another. An application of a best first search for the best subset identification is carried out to avoid searching through all possible attribute subsets.

\subsubsection{Wrapper}
A wrapper refers to a category of attribute selection techniques in which the significance of an attribute subset is estimated from the resulting classification accuracy achieved by a classifier~\citep{Kohavi}. In other words, the ability of a wrapper to identify necessary attributes or input features depends on the chosen classifier. Repeated five-fold cross-validation is implemented to provide an estimate of classification accuracy when an attribute subset is considered. Basically, the data samples are randomly divided into five folds where four folds of samples are used to train the classifier while the remaining fold of samples is used to test the classifier. The classifier training and testing procedure is carried out five times during one repetition of cross-validation where for each time a different sample fold is chosen as the testing fold. Hence, the samples in each fold are always used both to train and to test the classifier. Cross-validation is repeated as long as the standard deviation of classification accuracy over the repetitions is greater than one percent of the average classification accuracy or until the maximum of five repetitions is exhausted. The search for the best attribute subset is carried out via an application of a best first search and the chosen classifier for the wrapper is a na\"ive Bayes classifier.

\subsection{Na\"ive Bayes classifier}
A na\"ive Bayes classifier is a classification system in which the prediction of the class output is based on the application of Bayes theorem~\citep{Mitchell}. The na\"ive Bayes classifier can probabilistically predict the output class of an unknown sample using the available training samples to calculate the most probable output. The na\"ive Bayes classifier functions by assuming that the attribute values are conditionally independent given the output class. This assumption is particularly valid in this study because it is desirable to extract AIMs from different genomic regions, implying that the selected SNPs are most likely be uncorrelated.

\subsection{Implementation}
The round robin symmetrical uncertainty ranking and $F_{ST}$ extremity measurement programs are implemented in a C\# programming language. The programs have been successfully tested for the execution under Windows operating systems. On the other hand, the simple symmetrical uncertainty ranking, the correlation-based feature selection technique, the wrapper embedded with a na\"ive Bayes classifier and the na\"ive Bayes classifier are available as parts of a WEKA package~\citep{Witten}. All results included in the study are collected from the execution of the developed programs and WEKA in a personal computer. The computer is equipped with an Intel Core 2 Duo E6600 2.4 GHz processor and 2 GB of main memory. Windows XP is installed on the computer.

\begin{table}[!t]
\caption{The number of SNP data partitions from each chromosome.}
{\NINE
\begin{tabular*}{\textwidth}{@{\extracolsep{\fill}}lccccc@{}}
\toprule
          & {\it Number of SNP} & & {\it Number of SNP} & & {\it Number of SNP}\\      
{\it Chr} & {\it data partitions} & {\it Chr} & {\it data partitions} & {\it Chr} & {\it data partitions}\\            
\midrule        
        \phantom{0}1 & 59 & \phantom{0}9 & 35 & 16 & 21 \\
        \phantom{0}2 & 63 & 10 & 40 & 17 & 17 \\
        \phantom{0}3 & 48 & 11 & 38 & 18 & 23 \\
        \phantom{0}4 & 46 & 12 & 36 & 19 & 11 \\
        \phantom{0}5 & 47 & 13 & 30 & 20 & 23 \\
        \phantom{0}6 & 52 & 14 & 23 & 21 & 10 \\
        \phantom{0}7 & 41 & 15 & 20 & 22 & 11 \\ 
        \phantom{0}8 & 41 \\
\botrule
\end{tabular*}}
\label{tab:NumPartition}       
{\vskip2pt\NINE The number of partitions from Chromosome 2 is highest since there are more SNPs on this chromosome than other chromosomes. There are 735 partitions in total.} \vskip-12pt
\end{table}

\section{Results and discussion}
\subsection{Benchmarking of attribute selection techniques}
\begin{figure}[t!]
\centering
\caption{Performance of CFS, NB-Wrap, the simple SU ranking and the SU$_2$ ranking in conjunction with a na\"ive Bayes classifier.}
\includegraphics[width=12.5cm]{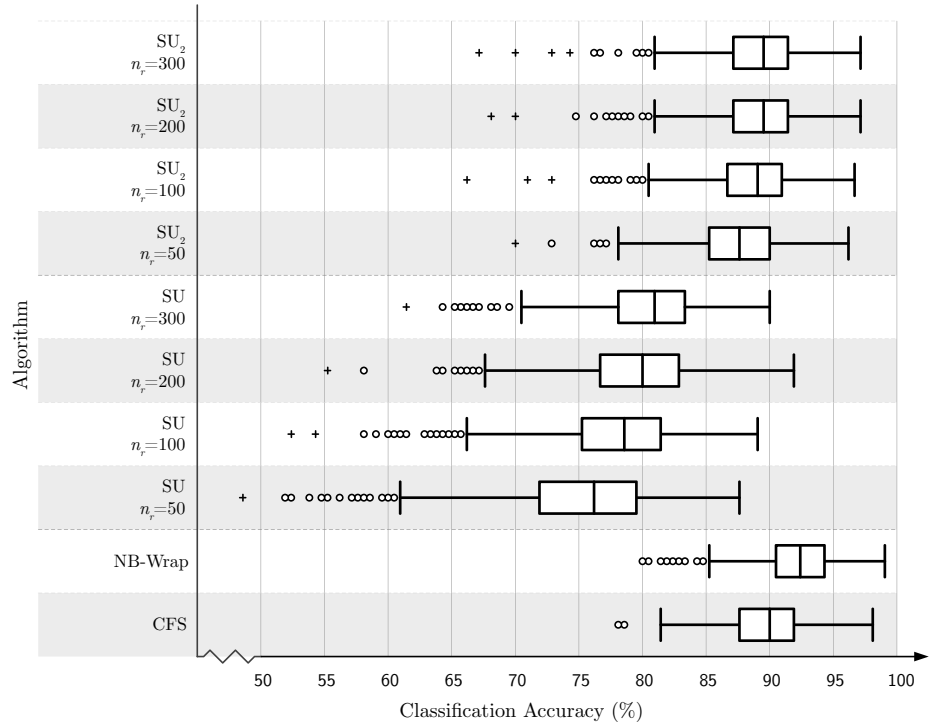}\\
\label{fig:fullcompare}
\end{figure}

There are many attribute selection techniques that can be used for AIM identification. A well-defined AIM panel should contain uncorrelated SNPs that lead to the highest population classification performance. Hence, a suitable attribute selection technique must be capable of extracting such an AIM panel from a SNP data set, which contains both correlated and uncorrelated SNPs. Furthermore, the computational time for the extraction process must be tractable. In this study, the candidate attribute selection techniques are the correlation-based feature selection (CFS) technique, the wrapper embedded with a na\"ive Bayes classifier (NB-Wrap), the simple symmetrical uncertainty ranking (simple SU ranking) and the newly proposed round robin symmetrical uncertainty ranking (SU$_2$ ranking). The classification performance is measured by applying selected attributes as inputs to a na\"ive Bayes classifier where ten-fold cross-validation is applied during the experiment. The values of $n_r$ (number of top-ranked SNPs) for both simple SU and SU$_2$ ranking techniques are set to 50, 100, 200 and 300. Since the HapMap data set contains a large amount of SNPs, the complete data set can be partitioned into a number of smaller data sets. Using multiple small data sets during the benchmarking of attribute selection techniques provides multiple results for statistical analysis. Moreover, it reduces the number of falsely selected attributes, which are unnecessary for the classification task, in each panel since the number of attributes available for selection in each data set is small.~\citep{Park}. Data partitioning is conducted on SNP data from every chromosome. Each partition covers on average 3,791,076 bases and consists of 5,000 positionally consecutive SNPs except for the last partition from each chromosome, which is allowed to contain less than 5,000 SNPs. The number of data partitions from each chromosome is summarised in Table~\ref{tab:NumPartition}. The search for AIMs is thus conducted by limiting the SNP inputs to those from the same partition. The distribution of classification accuracy obtained from all experiments is illustrated in Figure~\ref{fig:fullcompare}. It can be clearly seen that NB-Wrap produces the best screening result while CFS and the SU$_2$ ranking have the second best and third best results, respectively (a paired {\it t}-test on pair-wise algorithm comparison based on 735 experiments yields a {\it p}-value $<$ 0.05). The statistical power analysis also reveals that the benchmark trial with 735 data sets is sufficient for an accurate evaluation of overall algorithm performance (power $>$ 0.95 for a type I error rate of 0.05). It proves that the size of data partition is sufficient for the benchmark trial. These results can be further interpreted as follows.

%
%

Generally, attribute selection can significantly improve the classification efficacy. \cite{Hall} have performed a benchmark test on a number of attribute selection techniques. Similar to the results from the current study, wrappers and CFS are also proven to be the best and second best techniques, respectively. This is deduced from the overall classification performance across a range of benchmark problems in the UCI Machine Learning Repository in which the comparison is conducted by observing the performance of a na\"ive Bayes classifier before and after the attribute reduction. Wrappers and CFS appear to function well under moderate levels of attribute interaction. This is because both wrappers and CFS evaluate the significance of each attribute by considering the correlation between the attribute and the class while at the same time monitoring the inter-attribute correlation. Hence, a collection of attributes that together lead to high classification accuracy can often be conveniently identified.

The simple SU and SU$_2$ ranking techniques on the other hand consider only the correlation between each attribute and the class. Hence, the techniques are only able to identify the likelihood of an attribute being useful to the classification. In the absence of inter-attribute correlation monitoring, the presence of correlation among attributes can lead to performance degradation. The most probable source of attribute correlation is linkage disequilibrium, which exists among SNPs from the same localised region. The effect of linkage disequilibrium is most obvious when SNPs are screened by the simple SU ranking. \cite{Hastie} suggest that dividing a multi-class problem into a set of two-class problems can reduce the problem complexity. This approach has been adopted through the design and implementation of SU$_2$ ranking. By taking into account only a pair of populations at a time, each set of top-ranked SNPs generated prior to the set merging contains strong AIM candidates for separating two populations. Since a linkage disequilibrium pattern is specific to a population, a pattern difference is conveniently detectable when a pair-wise population comparison is conducted. This consequently leads to the reduction of linkage disequilibrium effect on the ranking mechanism as seen from the performance improvement exhibited by the SU$_2$ ranking over that from the simple SU ranking.

%
%
\begin{figure}[t!]
\centering
\caption{Performance of NB-Wrap and the two-stage approach, consisting of the SU$_2$ ranking and NB-Wrap, in conjunction with a na\"ive Bayes classifier.}
\includegraphics[width=12.5cm]{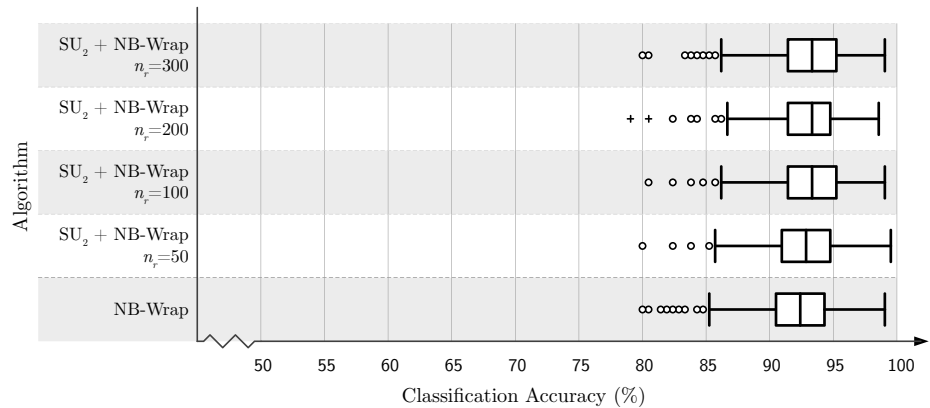}\\
\label{fig:partcompare}
\end{figure}

Although NB-Wrap produces the best screening result, its drawback is that a large computational effort is required to achieve this high performance. The computational time to finish the NB-Wrap calculation for each SNP partition on the computer is approximately 30 minutes while it takes less than one minute to complete the SU$_2$ ranking calculation. This is because the SU$_2$ ranking and NB-Wrap can tackle an $n$-attribute problem in linear and exponential time, respectively. Since the difference between the performance of both techniques is small, it is worth to explore the possibility of combining NB-Wrap and the SU$_2$ ranking. A similar two-stage approach for attribute selection has also been successfully applied in genetic association studies~\citep{Wongseree}. Basically, the SU$_2$ ranking is first applied to the data. The screened SNPs are subsequently used as inputs for NB-Wrap. The classification accuracy is hence determined from a na\"ive Bayes classifier that takes inputs from the finally screened SNPs. Ten-fold cross-validation is still employed during the experiment. The distribution of classification accuracy in Figure~\ref{fig:partcompare} suggests that the two-stage approach is capable of maintaining the same level of performance achieved by NB-Wrap regardless of the $n_r$ setting (a paired {\it t}-test on 735 experimental results yields a {\it p}-value $>$ 0.05). Furthermore, the two-stage approach with $n_r$ = 50, 100, 200 and 300  leads to a reduction of computational time from 30 minutes to two minutes. This proves that the two-stage approach is highly suitable for AIM identification. Hence, the two-stage approach is selected for the attribute selection step of the AIM identification protocol. Since the $n_r$ setting has no effect on the performance and the computational time of the two-stage approach, $n_r$ = 50 is the chosen setting for the application of the AIM identification protocol to the genome-wide data in the next section.

\subsection{Application of the AIM identification protocol to the HapMap data}
\begin{figure}[t!]
\centering
\caption{Empirical $F_{ST}$ distribution for every population pair. Among 3,619,209 SNPs in the HapMap data set, 1,303,591 loci are monomorphic in both CHB and JPT populations. Furthermore, the genotype at each locus is the same for both populations. As a result, it is not possible to calculate $F_{ST}$ values for the CHB-JPT population pair at these loci.}
\includegraphics[width=12.6cm]{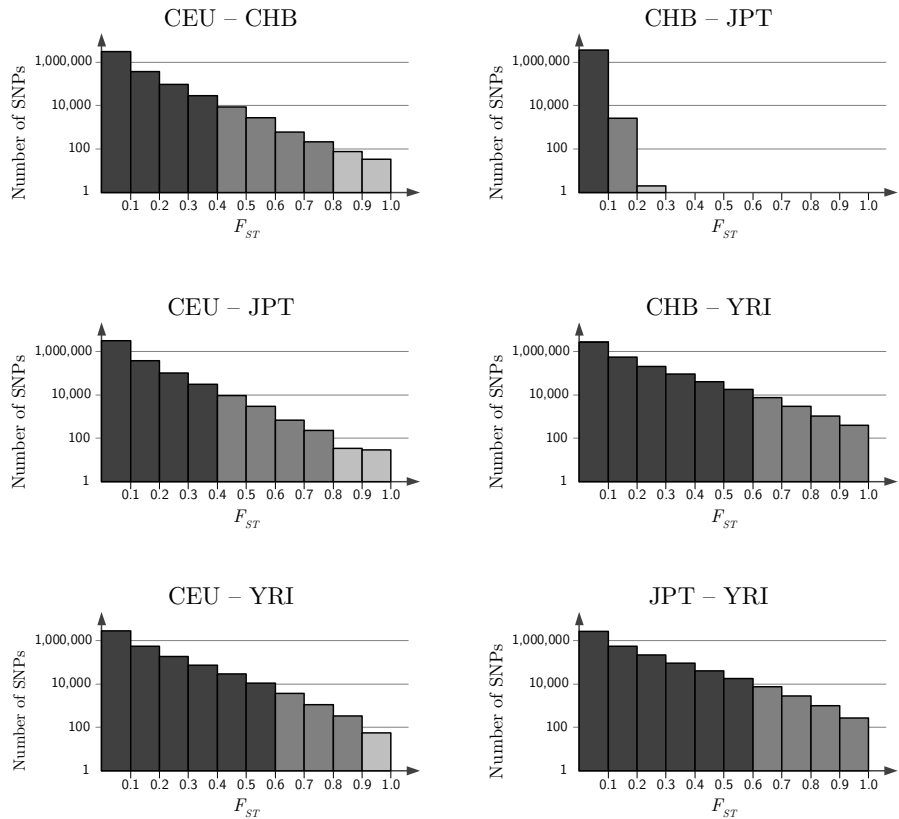}\\
\label{fig:Fdistribution}
\end{figure}

Candidate SNPs for inclusion in an AIM panel are SNPs from potential positive selection regions. These SNPs must have extreme $F_{ST}$ values in which the $F_{ST}$ extremity is estimated from empirical distribution. The empirical $F_{ST}$ distribution calculated for every population pair using all SNPs in the HapMap data is illustrated in Figure~\ref{fig:Fdistribution}. A similar $F_{ST}$ distribution calculated from the HapMap data has also been reported~\citep{Fujimoto}. The illustrated distribution describes different degrees of population subdivision for each population pair. For instance, the right tail of the $F_{ST}$ distribution for the CHB-JPT population comparison is located at a low numerical value, suggesting that these populations have recently begun to subdivide. On the other hand, the right tail of the $F_{ST}$ distribution for the CEU-YRI, CHB-YRI and JPT-YRI population pairs is situated at a high numerical value. This implies that the emergence of newer populations from the African ancestors has taken place a long time ago. There are 31,465 SNPs with $F_{ST}$ values in the top 0.3 percentile of the illustrated distribution. The cut-off of 0.3 percentile is derived from the mean added by six times of the standard deviation~\citep{Tennant}. Each SNP possesses at least one extreme $F_{ST}$ value among six $F_{ST}$ values obtained from the pair-wise population comparison. SNPs with extreme $F_{ST}$ values are subsequently screened by removing SNPs which are located neither inside nor close to genes. SNPs that are close to a gene are located within 2,000 bases upstream of the start site or downstream of the termination site for transcription. The resulting candidate SNP set contains 13,328 SNPs within or near genes. A non-synonymous SNP set is also derived from the full candidate set and contains 230 SNPs. The full candidate and non-synonymous SNP sets are then subjected to two-stage attribute selection---the SU$_2$ ranking with $n_r = 50$ follows by NB-Wrap---and na\"ive Bayes classification.

%
%

\begin{table}[!t]
\caption{Ten SNPs selected from the full candidate set containing 13,328 SNPs that have $F_{ST}$ values in the top 0.3 percentile of the empirical distribution and lie within or near genes.}
{\NINE
\begin{tabular*}{\textwidth}{@{\extracolsep{\fill}}lccccc@{}}
\toprule
          & {\it Population} &          & {\it SNP location} &            & {\it Gene location on} \\
{\it SNP} & {\it pair}       & $F_{ST}$ & {\it on the gene}  & {\it Gene} & {\it the chromosome} \\  
\midrule       
        rs17408457 & CHB-JPT & 0.0974 &	Intron 7\phantom{0} & {\it SLC30A7} & 1p21.2 \\
        rs922452   & CEU-CHB & 0.9804 &	Intron 4\phantom{0} & {\it EDAR}	& 2q11--q13 \\
	               & CEU-JPT & 0.7310 \\			
	               & CHB-YRI & 0.9063 \\			
	               & JPT-YRI & 0.6593 \\			
        rs12633912 & CHB-JPT & 0.1180 &	Intron 3\phantom{0} & {\it FOXP1}   & 3p14.1 \\
        rs2693740  & CEU-CHB & 1.0000 &	Intron 1\phantom{0} & {\it UBE2H}	& 7q32 \\
	               & CEU-JPT & 1.0000 \\			
	               & CHB-YRI & 1.0000 \\			
	               & JPT-YRI & 1.0000 \\			
        rs10758590 & CHB-JPT & 0.1704 &	Intron 2\phantom{0} & {\it GLIS3}   & 9p24.2 \\
        rs3803464  & CHB-JPT & 0.0806 &	Intron 8\phantom{0} & {\it MAN2C1}	& 15q11--q13 \\
        rs2238298  & CHB-JPT & 0.1100 &	Intron 18           & {\it POLG}	& 15q25 \\
        rs2526371  & CHB-JPT & 0.1353 &	Intron 2\phantom{0} & {\it RNF43}	& 17q22 \\
        rs6074677  & CHB-JPT & 0.1097 &	Intron 2\phantom{0} & {\it MACROD2} & 20p12.1 \\
        rs2269529  & CEU-YRI & 0.9355 &	Exon 34\phantom{0}  & {\it MYH9}    & 22q13.1 \\
\botrule
\end{tabular*}}
\label{tab:SNPinGene}       
\end{table} 

The panel of 10 SNPs from the full candidate set given in Table~\ref{tab:SNPinGene} leads to complete classification between four populations. There are two SNPs which could signify positive selection: rs922452 and rs2269529. rs922452 is located on intron 4 of {\it EDAR}. This SNP and rs3827760 in the CHB samples are in linkage disequilibrium ($D'$ $>$ 0.9). rs3827760 is a non-synonymous missense SNP which is located on exon 12 of {\it EDAR} and causes a conservative substitution of valine by alanine. rs3827760 has been subjected to many investigations and is proven to be the source of positive selection of {\it EDAR} in Asian populations~\citep{Sabeti07, Bryk, Fujimoto, Mou}. The discovery of an extreme $F_{ST}$ AIM in the proximity of rs3827760 conforms to the evidence that the average $F_{ST}$ value across the {\it EDAR} SNPs is significantly higher than the genome-wide average $F_{ST}$ value~\citep{Kelley}. In contrast to rs922452, rs2269529 is a non-synonymous missense SNP which is located on exon 34 of {\it MYH9} and causes a conservative substitution of isoleusine by valine. The genotype distribution at this polymorphic locus suggests that positive selection may have occurred in the CEU population since the ancestral allele A is entirely replaced by the derived allele G. Moreover, an analysis of the HapMap data suggests that {\it MYH9} is located in a low heterozygosity region~\citep{Cheng09}. Further studies on the effect of rs2269529 on possible genotypic changes are required to confirm that positive selection has in fact occurred.

\begin{table}[!t]
\caption{Sixteen SNPs selected from 230 non-synonymous SNPs with $F_{ST}$ values in the top 0.3 percentile of the empirical distribution.}
{\NINE
\begin{tabular*}{\textwidth}{@{\extracolsep{\fill}}lccccc@{}}
\toprule
          &                  &          & {\it Type of}        &            & {\it Gene location}\\
          & {\it Population} &          & {\it non-synonymous} &            & {\it on the} \\
{\it SNP} & {\it pair}       & $F_{ST}$ & {\it missense SNP}   & {\it Gene} & {\it chromosome} \\  
\midrule
        rs4915691 &	CEU-YRI & 0.5428 & Conservative & {\it DNAJC6} & 1pter--q31.3 \\
        rs3795661 &	CHB-JPT	& 0.0918 & Conservative	& {\it ZNF697} & 1p12 \\
        rs3827760 &	CEU-CHB	& 0.9247 & Conservative	& {\it EDAR}   & 2q11--q13 \\
	              & CEU-JPT	& 0.6917 \\			
	              & CHB-YRI & 0.9247 \\			
	              & JPT-YRI & 0.6917 \\			
        rs1366842 &	CHB-YRI	& 0.6732 & Non-conservative & {\it ZNF804A} & 2q32.1 \\
	              & JPT-YRI	& 0.8314 \\			
        rs482912  &	CHB-JPT	& 0.0790 & Conservative	& {\it LAMP3} &	3q26.3--q27 \\
        rs2294008 &	CHB-JPT	& 0.1700 & Non-conservative	& {\it PSCA} & 8q24.2 \\
        rs10989591 & CHB-JPT & 0.0932 &	Conservative & {\it	GRIN3A}	& 9q31.1 \\
        rs284859  &	CHB-JPT	& 0.1049 & Non-conservative & {\it C10orf26} & 10q24.32 \\
        rs6265	  & CHB-JPT	& 0.0864 & Conservative & {\it BDNF} & 11p13 \\
        rs735295  &	CHB-JPT	& 0.1055 & Non-conservative & {\it CCDC77} & 12p13.33 \\
        rs2228224 &	CHB-JPT & 0.0942 & Non-conservative	& {\it GLI1} & 12q13.2--q13.3 \\
        rs9262	  & CHB-JPT	& 0.1060 & Conservative & {\it C12orf29} & 12q21.32 \\
        rs2273801 &	CHB-JPT & 0.0806 & Conservative	& {\it WDR25} &	14q32.2 \\
        rs2337127 &	CHB-JPT	& 0.1142 & Non-conservative & {\it LOC100130736} & 15q13.1 \\
        rs2236695 &	CHB-JPT	& 0.0940 & Conservative	& {\it PRDM15} & 21q22.3 \\
        rs5764698 &	CHB-JPT	& 0.0942 & Conservative	& {\it SMC1B} &	22q13.31 \\
\botrule
\end{tabular*}}
\label{tab:MissenseSNP}       
\end{table}   

The panel of 16 SNPs from the non-synonymous SNP set given in Table~\ref{tab:MissenseSNP} also leads to complete classification between four populations. Unsurprisingly, rs3827760, which is located in {\it EDAR}, is present in the panel. The obtained $F_{ST}$ values support the presence of subdivision between CHB/JPT and CEU populations and that between CHB/JPT and YRI populations. This conforms to the evidence of positive selection of {\it EDAR} in Asian populations. In addition to rs3827760, rs1366842 is another SNP that indicates the subdivision between CHB and YRI populations and that between JPT and YRI populations. rs1366842 is located on exon 4 of {\it ZNF804A}, which is identified as a candidate gene for recent positive selection in Pima Indians~\citep{LopezHerraez}. Both rs3827760 and rs1366842 are the only two SNPs in the panel that are useful for separating CHB and JPT populations from the other populations. In contrast, only rs4915691 is required to identify the subdivision between CEU and YRI populations. Since the genotype distribution in the CEU population at rs4915691 is similar to that at rs2269529 on {\it MYH9}, rs2269529 is not needed in this AIM panel.

The identification of non-synonymous missense SNPs, which are also AIMs, provides a direct correlation between possible phenotypic variations and ancestral origins. These possible phenotypic variations are the results of both conservative and non-conservative substitutions of amino acids. A non-conservative missense SNP is more likely to produce noticeable consequences since it causes the substitution of an amino acid with different properties. Nonetheless, the phenotypic effect of a non-synonymous missense SNP remains difficult to predict since it depends on how the substitution of an amino acid changes the structure and function of a protein~\citep{Hartwell}.

Both AIM panels reported in this study lead to complete classification between four populations in the HapMap data. As a result, the AIM panels are suitable for a validation study involving the classification of a much larger number of individuals from all four populations. However, there are advantages and disadvantages of choosing each panel. In terms of biological explanation, the panel of 16 non-synonymous missense SNPs offers a direct hypothesis that could lead to the identification of phenotypic variations observable as evidence of population subdivision. Although the panel of 10 SNPs from the full candidate set also offers a similar hypothesis, the task of hypothesis testing is not straightforward since the majority of SNPs in the panel are intronic SNPs. In other words, linkage disequilibrium analysis is required to identify the root causes of phenotypic variations. Nevertheless, the reduction in genotyping cost attained by reducing the number of SNPs in the AIM panel from 16 to 10 may be significant since the number of individuals required for a validation study is usually large.

\subsection{Comparison with other AIM identification techniques}
In the present study, the genome-wide search for AIMs reveals that sets of 10 and 16 SNPs are sufficient for complete classification between four populations in the HapMap data. The sizes of AIM panels are at least four times smaller than those reported in the early works by \cite{Park}, \cite{Paschou} and \cite{Zhou}. A summary of the sizes of AIM panels from the early works and the present study is given in Table~\ref{tab:NumSNP}. \cite{Park} employ a nearest shrunken centroid method while \cite{Zhou} develop a modified {\it t}-test for SNP screening. Both approaches are filter-based attribute selection techniques where each SNP is prioritised by identifying its usefulness for separating all population classes from one another. This is different from the strategy embedded in the SU$_2$ ranking in which each SNP is prioritised according to its usefulness for separating classes in each class pair. This strategic difference is most likely to be the cause of the reduction in the sizes of AIM panels from those reported in the works by \cite{Park} and \cite{Zhou}. Nevertheless, the strategy employed in the SU$_2$ ranking can be incorporated into both the nearest shrunken centroid method and the modified {\it t}-test. The modification should enhance the capability of both approaches, which could lead to the reduction in the sizes of AIM panels. In contrast to \cite{Park} and \cite{Zhou}, \cite{Paschou} use a clustering technique to identify AIMs. In other words, the population labels are not considered during the SNP screening. As a result, larger AIM panels than those from the present study are selected to achieve the maximum distances between population clusters. Although the technique proposed by \cite{Paschou} may be less effective in the case of HapMap data, the technique is necessary when the population labels are not known a priori and the population boundary is determined solely via genetics.

\begin{table}[!t]
\caption{Number of SNPs required for the classification of HapMap data.}
{\NINE
\begin{tabular*}{\textwidth}{@{\extracolsep{\fill}}lccc@{}}
\toprule
                        & {\it Number of} &  {\it Number of} & {\it Classification} \\    
{\it Reference} & {\it populations} & {\it SNPs}      & {\it accuracy (\%)} \\       
\midrule         
Present study & 4 & \phantom{0}10 & 100.00 \\
                      & 4 & \phantom{0}16 & 100.00 \\
\cite{Park}      & 3 & \phantom{0}82 & 100.00 \\
\cite{Zhou}     & 3 & \phantom{0}64 & 100.00 \\
                      & 4 & 100 & \phantom{0}90.00 \\
\cite{Paschou} & 3 & \phantom{0}14 & 100.00 \\
                       & 4 & 164 & \phantom{0}99.52 \\
	              & 4 & \phantom{0}64 & \phantom{0}98.57 \\
\botrule
\end{tabular*}}
\label{tab:NumSNP}       
{\vskip2pt\NINE	The three-population problem is formulated by grouping JPT and CHB samples into the same class. \cite{Paschou} report two AIM panels, which are identified using different settings of the desired panel size, for the four-population problem. With the use of 64 SNPs one CHB sample and two JPT samples are misclassified, while with the use of 164 SNPs only one JPT sample is misclassified. Two AIM panels for the four-population problem are also reported in the present study. The panel of 10 SNPs is extracted from a full candidate set which contains 13,328 SNPs that have extreme $F_{ST}$ values and lie within or near genes. On the other hand, the panel of 16 SNPs is extracted from a non-synonymous SNP set which contains 230 SNPs with extreme $F_{ST}$ values.} \vskip-12pt
\end{table}

Generally, AIMs are used to infer ancestry proportions of individuals in an admixed population in which the total number of ancestral populations or classes and the ancestral population label of each individual are not known beforehand. In such a case, it is not possible to directly apply the proposed AIM identification protocol to the problem. The population structure must first be identified where the appropriate number of classes is chosen and ancestral population labels are assigned to all individuals~\citep{Pritchard, Tang, Paschou}. Then the proposed protocol can be applied to identify SNPs necessary for population classification. Moreover, it is anticipated that the proposed protocol would return larger AIM panels than those reported for four populations in the HapMap data. This is because each individual in an admixed population could inherit SNPs from multiple ancestral populations. In other words, the population boundaries are not as distinctive as those between four populations in the HapMap data where individuals from each population are descendants of ancestors who are exclusive to that particular population.

In the present study, the predicted output class is obtained from a na\"ive Bayes classifier and is the class with the highest probability. This means that the proposed AIM identification protocol in its present form is not capable of identifying each individual from an admixed population as a member of multiple ancestral classses. Nonetheless, the proposed protocol can be modified to accommodate this scenario by reporting each probability value for selecting an ancestral class in the problem as the output instead of reporting only the output class with the highest probability. However, an additional criterion for determining the classification accuracy is also required.

The proposed AIM identification protocol relies on the ability to estimate $F_{ST}$ extremity of each SNP in the data set. If the available SNPs do not cover enough genomic regions, the empirical $F_{ST}$ distribution may significantly depart from the actual distribution. In addition to the constraint imposed by the $F_{ST}$ extremity estimation, the number of populations in the classification problem also places a limitation on the functionality of the proposed protocol. This is because $SU_2$ values for each SNP are calculated for every pair-wise population comparison during the SU$_2$ ranking. Nevertheless, the computational time of SU$_2$ ranking is a quadratic function of the number of populations. This means that the computational time is still tractable for a reasonably large problem.

\subsection{AIM transferability}
As mentioned earlier in the Introduction, SNPs from potential positive selection regions detected in a set of populations are selected as AIMs because they have a strong potential for being applicable as AIMs for the inference of other related populations. To demonstrate this rs2269529, which is located in {\it MYH9}, is chosen as the attribute for classification between ASW (African ancestry in Southwest USA) and CEU populations. Forty-eight unrelated ASW samples are obtained from the public release \#28 of HapMap data set (Phase III, release date: August 2010), which is available in NCBI build 36 (dbSNP b126) coordinates. The ASW population differs from the YRI population since the ASW population is affected by a significant non-African influence. Only one SNP is chosen for the demonstration because many SNPs required in the AIM panels reported in this study are not available in the ASW samples. In particular, there is no genotypic information for rs3827760 ({\it EDAR}), rs922452 ({\it EDAR}), rs1366842 ({\it ZNF804A}) and rs2693740 ({\it UBE2H}). These SNPs are crucial for the identification of CHB and JPT populations. Without the information, it is no longer possible to differentiate between CHB/JPT and ASW populations. With the use of rs2269529 as the only input for a na\"ive Bayes classifier, complete classification between ASW and CEU populations is achieved. The subdivision between ASW and CEU populations is confirmed by the $F_{ST}$ value of 0.8760. Although the result conforms to the motivation given above, there is no guarantee that complete classification between CEU and other African populations can still be attained. This is because other African populations could be affected by a stronger non-African influence than that experienced by the ASW population. In such a case, the classification accuracy would certainly reduce. Subsequently, the use of additional SNPs would be required to regain complete classification and hence leads to an increase in AIM panel size. The same caution is also applied to the attempt to generalise the AIM panels reported in this study in other classification problems that involve populations related to the CEU, CHB and JPT populations.

\section{Conclusion}
In this article, the identification of ancestry informative markers (AIMs) within potential positive selection regions has been conducted. The AIM identification protocol consists of three main steps: identification of SNPs with extreme $F_{ST}$ values, SNP screening via attribute selection and classification model construction. SNPs are primarily screened according to their $F_{ST}$ values. The $F_{ST}$ extremity is estimated from the empirical $F_{ST}$ distribution evaluated from all SNPs in the genome-wide data. SNPs with extreme $F_{ST}$ values are subjected to further screening by two-stage attribute selection consisting of round robin symmetrical uncertainty ranking and a wrapper embedded with a na\"ive Bayes classifier. Finally, a classification model is built from the finally screened SNPs using a na\"ive Bayes classifier. Ten-fold cross-validation is applied during the AIM search. The proposed protocol is implemented and tested on the HapMap Phase II data set, which covers samples from four populations namely the CEU, CHB, JPT and YRI populations~\citep{HapMap03, HapMap05, HapMap07}. Two AIM panels are identified. The first panel containing 10 SNPs is extracted from a candidate set of SNPs located within or near genes while the second panel containing 16 SNPs is extracted from a non-synonymous SNP set. Both panels are made up from lesser numbers of SNPs than those previously reported~\citep{Park, Paschou, Zhou}. This suggests that a synergy between information extracted by data mining and that based on prior knowledge regarding population subdivision and locations of genic SNPs leads to more efficient AIM identification. The limitation of the proposed protocol and how it can be improved are also discussed.

\section*{Acknowledgements}
The authors acknowledge three anonymous reviewers, Prof. Taesung Park and Prof. Xiaohua (Tony) Hu for their valuable comments and suggestions on this article. Damrongrit Setsirichok, Theera Piroonratana and Waranyu Wongseree were supported by the Thailand Research Fund (TRF) through the Royal Golden Jubilee Ph.D. Programme (Grant No. PHD/1.E.KN.51/A.1, PHD/1.E.KN.50/A.1 and PHD/1.E.KN.49/A.1, respectively). Touchpong Usavanarong was supported by the Faculty of Engineering of the King Mongkut's University of Technology North Bangkok. Chanin Limwongse was supported by the Mahidol Research Grant. Nachol Chaiyaratana was supported by the Thailand Research Fund, the Office of the Higher Education Commission and the Faculty of Engineering of the King Mongkut's University of Technology North Bangkok.

\end{document}